\documentstyle[epsf]{l-aa}
\begin{document}
\thesaurus{02.04.2; 06.01.1; 06.05.1; 06.09.1; 06.18.2}
\title{Local Mixing near the Solar Core, Neutrino Fluxes and
Helioseismology}

\author{Olivier Richard \and Sylvie Vauclair}

\institute{
 Observatoire Midi-Pyr\'en\'ees
   14, avenue Edouard Belin, 31400 Toulouse, France}

\offprints{Sylvie Vauclair}

\date{Received ; accepted }

\maketitle

\markboth{Olivier Richard and Sylvie Vauclair}{Local Mixing near the Solar Core, Neutrino Fluxes and
Helioseismology}

\begin{abstract}
We have computed solar models similar to those published in Richard, Vauclair, Charbonnel, Dziembowski (1996), in which we have added local mixing near the solar core in order to decrease the neutrino production. The results show that the neutrino fluxes are reduced as expected (although not enough to account for the observed values), but the obtained models are incompatible with the inversion of the helioseismic modes. We have specially tested the parametrized mixing proposed by Morel and Schatzman (1996). The resulting solar models are far from the seismic model and thus unrealistic.
\end{abstract}
We have computed new solar models using the same stellar 
evolution code as described in Charbonnel, Vauclair and 
Zahn (1992). This code, originating from Geneva, now 
includes the computation of element segregation for helium 
and 12 heavier isotopes. It may also include any type of 
mixing of the stellar gas, provided this mixing may be 
parametrised with an effective diffusion coefficient as a 
function of radius.

In the present computations, we have introduced in the  
solar model 4 of Richard et al (1996) (hereafter model T1)
a parametrized mixing region
located at the edge of the nuclear burning core. Such a mixing, 
which could be induced by Stochastic Internal Waves (Morel 
and Schatzman 1996), remains as an a priori possibility to 
decrease the solar neutrino flux. The basic reason is that it 
brings $\displaystyle ^{3}$He down towards the solar 
center (figure~1) and increases the rate of the $\displaystyle 
^{3}$He ($\displaystyle ^{3}$He, 2p)~$\displaystyle ^{4}$He 
nuclear reaction yield, while the 
$\displaystyle ^{3}$He~($\displaystyle ^{4}$He,~$\displaystyle \alpha
$)~$\displaystyle ^{7}$Be 
reaction 
is reduced.

We have introduced this extra-mixing in the form of a 
gaussian, of the type:
$$
D = D_{0} \exp \left[- \left({r - r_{c} \over 2\Delta }
 \right) ^{2}\right]
$$
The comparisons of the $\displaystyle u = P / \rho $
 function in these models and in the seismic Sun as a
function of radius are  unsatisfactory. The
very good agreement obtained by Richard et al (1996) for the
models including microscopic diffusion and a mild mixing
below the convection zone is  destroyed in most
models which include the core mixing.
It is possible to keep the good agreement in the outer parts
of the Sun
while perturbating only the central regions: for this a
``cut-off'' of the mixing
effect must be introduced
at a fractional radius of
$\displaystyle r/R_{\odot} = .4$. However, even in
this case, the $ u $ values in the core are incompatible with the
seismic Sun (figure~2).

Here we show the results obtained for two different models, 
model T2 (similar to model 20431 in Morel
and Schatzman 1996) with the following values for the gaussian parameters:

$\displaystyle {r_{c} /R_{\odot}} = $ 
.2 \hspace{.5cm} 
$\displaystyle D_{0} = $\, 1000 
~cm$\displaystyle ^{2}$s$\displaystyle ^{-1}$
\hspace{.5cm} 
 $\displaystyle \Delta /R_{\odot} = $ .04

and model T3 with the following values for the gaussian parameters:

$\displaystyle {r_{c} /R_{\odot}} = $ 
.15 
\hspace{.5cm}
$\displaystyle D_{0} = $\, 100 
~cm$\displaystyle ^{2}$s$\displaystyle ^{-1}$
\hspace{.5cm}
 $\displaystyle \Delta /R_{\odot} = $ .025

with a cut-off of the gaussian function at ${r /R_{\odot}} = .4$
The main physical parameters of these models are given in table 1.

In all cases we obtain a  decrease of the neutrino 
fluxes, although they remain too large to be compatible with the 
detection values (table~2 and 3).
The comparison with the helioseismological results show however that
these models are not realistic (figure 2).

In conclusion, although some local mixing inside the Sun 
may help reducing the neutrino fluxes, it cannot be 
reconciled with helioseismology. The helioseismic data 
prove to be a very powerful tool in constraining the 
remaining parameters of the solar structure. It will improve 
even more in the central parts when observations of gravity 
waves will be possible.

\begin{table}
\caption{\small{Main physical parameters of the three models, at the base of
the convective zone and at the center.
}}
\begin{center}
\begin{tabular}{cccccccc}
\hline
\multicolumn{1}{c}{model}
& \multicolumn{1}{c}{${r_{cz} /{R_{\odot}}}$}
& \multicolumn{1}{c}{T$_{cz}$}
& \multicolumn{1}{c}{$\rho_{cz}$}
& \multicolumn{1}{c}{Y$_c$}
& \multicolumn{1}{c}{X$_c$}
& \multicolumn{1}{c}{T$_c$}
& \multicolumn{1}{c}{$\rho_c$}\\
 & & (10$^6$K) & (g.cm$^{-3}$) & & &(10$^6$K) & (g.cm$^{-3}$)
\\ 
\hline
 T1 & 0.717 & 2.162 & 0.185 & 0.6431 & 0.3368 & 15.63 & 154.17 \\
 T2 & 0.724 & 2.060 & 0.156 & 0.5687 & 0.4113 & 14.96 & 127.05 \\
 T3 & 0.717 & 2.151 & 0.181 & 0.6337 & 0.3462 & 15.45 & 148.93 \\
\hline
\end{tabular}
\end{center}
\caption{Gaussian parameters and neutrino fluxes for the three models.}
\begin{center}
\begin{tabular}{ccccccc}
\hline
\multicolumn{1}{c}{model}
& \multicolumn{1}{c}{${r_{c} /{R_{\odot}}}$}
& \multicolumn{1}{c}{$ \Delta/R_{\odot} $}
& \multicolumn{1}{c}{$D_{0}$}
& \multicolumn{1}{c}{$\phi (^{8}$B)}
& \multicolumn{1}{c}{$(\phi \sigma )$Cl}
& \multicolumn{1}{c}{$(\phi \sigma )$Ga}\\
 & & & (cm$^{2}$s$^{-1}$) & 
(10$^{6}$ cm$^{2}$ s$^{-1}$) & (SNU$_{S}$) & (SNU$_{S}$) \\
\hline
 T1 &  -  &  -  & - & 6.06 & 8.14 & 130.84\\
 T2 & .20 & .040 & 1000 & 2.60 & 3.90 & 108.75\\
 T3 & .15 & .025 & 100 & 3.85 & 5.45 & 115.80\\
\hline
\end{tabular}
\end{center}
\caption{Detected values of the solar neutrino fluxes
("VIIIth Rencontres de Blois", in press).}
\begin{center}
\begin{tabular}{llll}
\multicolumn{1}{l}{}
&\multicolumn{1}{l}{}
&\multicolumn{1}{l}{}
& \multicolumn{1}{l}{}\\
  & $\phi (^{8}$B)& =& 2.80 $\pm$ 0.19 $\pm$  0.33 $\times 
10^{6}$ cm$^{2}$ s$^{-1}$\\ 
  & $ (\phi \sigma )_{\rm Cl}$& =& 2.54 $\pm$ 0.14 $\pm$ 0.14 
SNU$_{S}$\\
 Sage: & $(\phi \sigma )_{\rm Ga} $& =& 72$^{+12 + 5}_{-10 -7}$
 SNU$_{S}$\\
 Gallex:& $(\phi \sigma )_{\rm Ga} $& =& 69.7 $\pm$ 6.7$^{+ 3.9}_{- 4.5}$
SNU$_{S}$\\
\end{tabular}
\end{center}
\end{table}

\begin{figure*}
\begin{center}
\epsfxsize=16cm
\epsfysize=16cm
\epsfbox{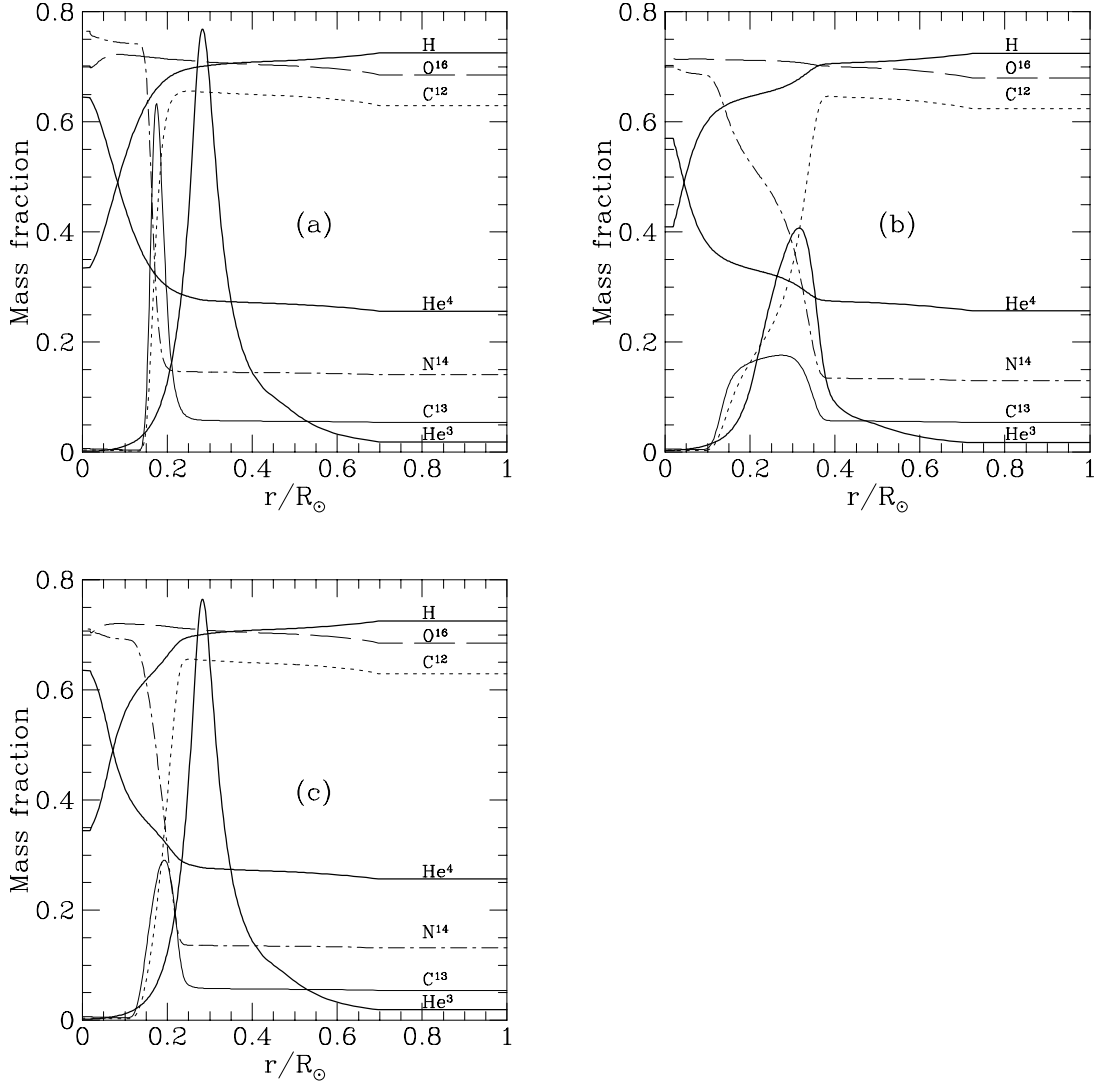}
\end{center}
\caption{
Abundance profiles for some 
elements in the three models (The mass fractions are  
multiplied by 250 for $ ^{3}$He, by 200 for
$^{12}$C, by 100 for $ ^{13}$C,
by 140 for $ ^{14}$N and by 75 for 
$ ^{16}$O):
graph (a) for model T1;  
graph (b) for model T2;
graph (c) for model T3.
}
\end{figure*}
\begin{figure*}
\begin{center}
\epsfxsize=15.5cm
\epsfysize=8cm
\epsfbox{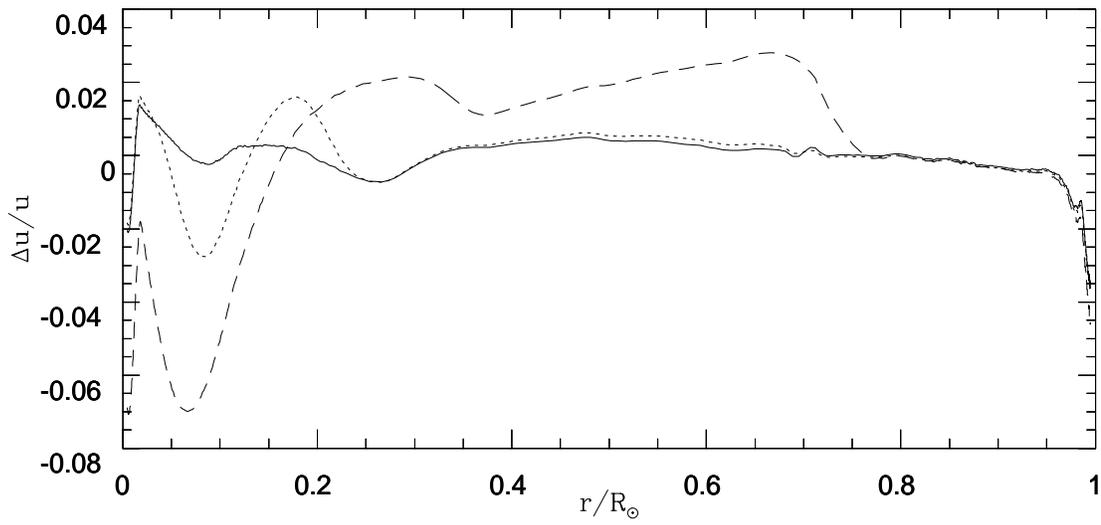}
\end{center}
\caption{
Difference between 
the $\left(u  = {P \over \rho } \right) $ 
deduced  from helioseismology and the computed one. The 
ordinates represents: $ {\Delta u \over u} =
{u \ \hbox{(seismic)} - u \ \hbox{(model)} 
\over u \ \hbox{(seismic)}  }$.
Solid line: model T1; 
dashed line: model T2
; dotted line: model T3.
These two last models, which include local mixing near the solar core,
are not compatible with the helioseismic results.}
\end{figure*}


\begin{thebibliography}{}
\bibitem[]{}
Charbonnel, C., Vauclair, S., Zahn, J.P., 1992, A\&A 255, 191
\bibitem[]{}
Morel, P., Schatzman, E., 1996, A\&A 310, 982
\bibitem[]{}
Richard, O., Vauclair, S., Charbonnel, C., Dziembowski,
W.A., 1996, A\&A 312, 1000
\end{thebibliography}
\end{document}